\documentclass[conference]{IEEEtran}
\usepackage{color}
\usepackage[english]{babel}
\usepackage{graphicx}
\usepackage{epstopdf}
\usepackage{times}
\usepackage{multirow}
\usepackage[algo2e,linesnumbered]{algorithm2e} 
\usepackage{algorithmic}
\usepackage{algorithm}
\usepackage{mathenv}
\usepackage{array}
\usepackage[T1]{fontenc}
\usepackage{array}
\usepackage{mdwmath}
\usepackage{multicol}
\usepackage{amsmath}
\usepackage{amssymb}
\usepackage{float}
\usepackage{url}
\newcommand{\subparagraph}{}
\usepackage{lipsum}
\usepackage{titlesec}
\titlespacing\section{0pt}{6pt plus 2pt minus 2pt}{6pt plus 2pt minus 2pt}
\titlespacing\subsection{0pt}{3pt plus 2pt minus 2pt}{3pt plus 2pt minus 2pt}

\usepackage{mathtools}
\DeclarePairedDelimiter\ceil{\lceil}{\rceil}
\DeclarePairedDelimiter\floor{\lfloor}{\rfloor}
\usepackage{dirtytalk}
\usepackage[utf8]{inputenc}
\usepackage{amsthm}
\newtheorem{theorem}{Theorem}
\newtheorem{definition}{Definition}
\makeatletter
\newcommand*{\rom}[1]{\expandafter\@slowromancap\romannumeral #1@}
\makeatother

\showoutput
\begin{document}
\long\def\/*#1*/{}
\setlength{\abovedisplayskip}{1pt}
\setlength{\belowdisplayskip}{1pt}

\title{Distributed Scheme for Interference Mitigation of Coexisting WBANs Using Latin Rectangles}

  \author{\IEEEauthorblockN{Mohamad Ali\IEEEauthorrefmark{1}, Hassine Moungla\IEEEauthorrefmark{1}, Mohamed Younis\IEEEauthorrefmark{2}, Ahmed Mehaoua\IEEEauthorrefmark{1}}
\IEEEauthorblockA{\IEEEauthorrefmark{1}LIPADE, University of Paris Descartes, Sorbonne Paris Cit\'{e}, Paris, France \\
}{\IEEEauthorrefmark{2}Department of Computer Science and Electrical Engineering, University of Maryland, Baltimore County, United States} \\Email: \{mohamad.ali; hassine.moungla; ahmed.mehaoua\}@parisdescartes.fr; younis@umbc.edu}
\maketitle
\begin{abstract}
The performance of wireless body area networks (\textit{WBAN}s) may be degraded due to co-channel interference, i.e., when sensors of different coexisting \textit{WBAN}s transmit at the same time-slots using the same channel. In this paper, we exploit the \textit{16 channels} available in the \textit{2.4 GHz} unlicensed international band of \textit{ZIGBEE}, and propose a \underline{d}istributed scheme that opts to \underline{a}void \underline{i}nterference through channel to time-slot hopping based on \underline{L}atin rectangles, \textit{DAIL}. In \textit{DAIL}, each \textit{WBAN}'s coordinator picks a Latin rectangle whose rows are \textit{ZIGBEE} channels and columns are time-slots of its superframe. Subsequently, it assigns a unique symbol to each sensor; this latter forms a transmission pattern according to distinct positions of its symbol in the rectangle, such that collisions among different transmissions of coexisting \textit{WBAN}s are minimized. We further present an analytical model that derives bounds on the collision probability of each sensor's transmission in the network. In addition, the efficiency of \textit{DAIL} in interference mitigation has been validated by simulations.
\end{abstract}
\section{Introduction}
A \textit{WBAN} is a wireless short range communication network formed of a coordinator denoted by \textit{Crd} and multiple low power and miniaturized sensors that are placed inside or attached to the human body. These sensors collect health related data through continual monitoring of the physiological state of the body, while, a person is sitting, walking, running, etc. \textit{WBAN}s are used in various applications such as ubiquitous health care, sports and military \cite{key15}. For example, these sensors may be observing the heart (electrocardiography) and the brain electrical (electroencephalographs) activities as well as vital signs and parameters like insulin percentage in blood, blood pressure, temperature, etc.

Recently, the IEEE 802.15.6 standard \cite{key26} has proposed new specifications for \textit{WBAN}s that require the system to function properly within the transmission range of up to 3 meters when up to 10 \textit{WBAN}s are collocated. It also has to support 60 sensors in a $6m^{3}$ space (256 sensors in a $3m^{3}$). Thus, there is great possibility of interference amongst the collocated \textit{WBAN}s, e.g., in crowded areas such as a hospital lobby or corridor. Consequently, the interference may affect the communication links and degrade the performance of each individual \textit{WBAN}. Therefore, interference mitigation is of the utmost importance to improve the reliability of the whole network. To this end, the IEEE standard proposes three mechanisms for co-channel interference mitigation in \textit{WBAN}s, namely, beacon shifting, channel hopping and active superframe interleaving.

In addition, the co-channel interference is challenging due to the highly mobile and resource constrained nature of \textit{WBAN}s. Firstly, such nature makes the allocation of a global \textit{Crd} to manage multiple \textit{WBAN}s as well as the application of advanced antenna and power control techniques used in other networks unsuitable for \textit{WBAN}s. Secondly, due to the absence of coordination and synchronization among \textit{WBAN}s, the different superframes may overlap and the concurrent transmissions of different nearby \textit{WBAN}s may interfere. More specifically, when two or more sensors of different \textit{WBAN}s access the shared channel at the same time, their transmissions cause medium access collision. In this paper, we tackle these issues and contribute the following:
\begin{itemize}
 \item  \textit{DAIL, a distributed scheme that enables predictable time-based channel hopping using Latin rectangles in order to avoid interference among coexisting \textit{WBAN}s}
 \item \textit{An analytical model that derives bounds on the collision probability for sensors transmissions}
\end{itemize}
 The simulation results and theoretical analysis show that our proposed approach can significantly lower the number of collisions among the individual sensors of coexisting \textit{WBAN}s as well as increase the power savings at both \textit{sensor}- and \textit{WBAN}-levels. Moreover, \textit{DAIL} significantly avoids the inter-\textit{WBAN} interference and do not require any mutual coordination among the individual \textit{Crd}s. The rest of the paper is organized as follows. Section \rom{2} sets our work apart from other approaches in the literature. Section \rom{3} summarizes the system model and provides a brief overview of Latin squares. Section \rom{4} describes \textit{DAIL} in detail. Section \rom{5} analyzes the performance of \textit{DAIL}. Section \rom{6} presents the simulation results. Finally, the paper is concluded in Section \rom{6}.
\section{Related Work}
The problem of interference due to \textit{WBAN}s coexistence has been addressed through spectrum allocation, cooperation, power control, game theory and multiple access schemes. Example schemes that pursue the spectrum allocation methodology include \cite{key16, key25, key100, key200}. In \cite{key16}, a distributed spectrum allocation is proposed where inter-\textit{WBAN}s coordination is considered, the interfering sensors belonging to each pair of \textit{WBAN}s are assigned orthogonal sub-channels. Whereas, Movassaghi et al., \cite{key25} have proposed an adaptive interference mitigation scheme that operates on different parameters (e.g., nodes' traffic priority, signal strength, etc,) and allocates synchronous and parallel transmission intervals for interference avoidance. The proposed scheme has considered sensor-level interference for inter-network interference mitigation rather than considering each \textit{WBAN} as a whole. In the proposed \textit{DAIL} scheme, we have considered the interference at both sensor- and time-slot-levels. Meanwhile, in \cite{key100}, a prediction algorithm for dynamic channel allocation is proposed where, variations in channel assignment due to body gesture movements are factored in. The interference is avoided due to the allocation of transmission time based on synchronised clocks. It is worth noting that in \cite{key200} Latin squares are used in cellular networks for the sub-carrier allocations to users where, a user could be allocated multiple virtual channels. Each virtual channel hops over different sub-carriers at different orthogonal frequency division multiplexing (OFDM) symbol times. Basically, users are allocated multiple sub-carrier-to-OFDM-symbol-time combinations to avoid inter-cell interference. \textit{DAIL} sensors assigns single-channel-to-time-slot combination which simplifies inter-\textit{WBAN} coordination and time synchronisation.

A number of approaches have adopted cooperative communication, game theory and power control to mitigate co-channel interference. Dong et al., \cite{key9} have pursued joint a cooperative communication integrated with transmit power control based on simple channel predication for \textit{WBAN}s coexistence problem. Similarly, in \cite{key54}, a co-channel interference is mitigated using cooperative orthogonal channels and a contention window extension mechanism. 
Whereas, the approach of \cite{key30} employs a Bayesian game based power control to mitigate inter-\textit{WBAN} interference by modelling \textit{WBAN}s as players and active links as types of players . 

Other approaches pursued multiple access schemes for interference mitigation. Kim et al., \cite{key21} have proposed a distributed \textit{TDMA}-based beacon interval shifting scheme where, the wake up period of a \textit{WBAN} is made to not overlap with other \textit{WBAN}s by employing carrier sense before a beacon transmission. Whilst, Chen et al., \cite{key3} adopts \textit{TDMA} for scheduling transmissions within a \textit{WBAN} and carrier sensing mechanism to deal with inter-\textit{WBAN} interference. In \cite{key300}, many topology-dependent transmission scheduling algorithms have been proposed to minimize the \textit{TDMA} frame length in multihop packet radio networks using \textit{Galois field theory} and \textit{Latin squares}. For single-channel networks, the \textit{modified Galois field design} and the \textit{Latin square design} for topology-transparent broadcast scheduling is proposed. The \textit{modified Galois field design} obtains much smaller \textit{minimum frame length} than the existing scheme while the \textit{Latin square design} can even achieve possible performance gain when compared with the \textit{modified Galois field design}. In one-hop rather than multi-hop communication scheme, like \textit{DAIL}, using Latin squares better schedules the medium access and consequently significantly diminishes the inter-\textit{WBAN} interference.

In this paper, we take a step forward and exploit the \textit{16} channels available in the \textit{2.4 GHz} unlicensed international band (\textit{ISM}) of \textit{ZIGBEE} and, propose a distributed scheme based on channel and time-slot hopping for interference avoidance amongst coexisting \textit{WBAN}s. In our proposed \textit{DAIL} scheme, each \textit{WBAN} autonomously picks a Latin rectangle whose rows are the \textit{ZIGBEE} channels and columns are the time-slots that relates each channel to a time-slot within the Latin rectangle. Meanwhile, we depend on the special properties of Latin rectangles to minimize the probability of both time and channel matching among sensors in different \textit{WBAN}s.

\section{System Model and Preliminaries}
\subsection{System Model and Assumptions}
We consider the realistic scenario when \textit{N} \textit{TDMA}-based \textit{WBAN}s coexist in a crowded environment, e.g., when a group of patients moving around in a large hall of a hospital. Each \textit{WBAN} consists of a single \textit{Crd} and up to \textit{L} sensors, each denoted by \textit{SR} and generates its data based on a predefined sampling rate and transmits data at maximum rate of \textit{250Kb/s} within the \textit{2.4 GHz} \textit{ISM} band. Furthermore, we assume all \textit{Crd}s are equipped with unconstrained energy supply, e.g., equipped with harvesters, and are not affected by channel hopping.

Due to the \textit{WBAN}'s irregular and unpredictable motion pattern, it is very hard to achieve inter-\textit{WBAN}s coordination or to have a central unit to mitigate the potential interference when some of them are in proximity of each other. Basically, co-channel interference may arise due to the collisions amongst the concurrent transmissions made by sensors in different \textit{WBAN}s in the same time-slot denoted by \textit{Slt}. To address this issue, we exploit the \textit{16} channels in the \textit{ISM} band of \textit{ZIGBEE} to resolve this problem through combining the frequency with time hopping. \textbf{Table \ref{not}} summarizes the notations that we use.

\begin{table}
\centering
\caption{Notation Table}
\label{not}
\begin{tabular}{llll}
\noalign{\smallskip}\hline
\hline\noalign{\smallskip}
\textbf{Symbol}&\textbf{Deascription} &\textbf{Symbol} &\textbf{Deascription} \\
\hline
OLS& orthogonal Latin set&Crd&coordinator\\
P& transmission pattern&SP&superferame\\
M &\# of ZIGBEE channels&C&channel \\
L&\# of sensors per WBAN&Slt&time-slot\\
Q& max. \# of WBAN interferers&FL&frame length \\
K&\# of time-slots per SP& SR& sensor\\
ISM&international, scientific, medical&&\\
\noalign{\smallskip}\hline
\hline\noalign{\smallskip}
\end{tabular}
\end{table}

\subsection{Latin Squares}
In this section, we provide a brief overview of Latin squares that we used in our interference mitigation approach \cite{key300}.
\begin{definition}
A Latin square is a $\textit{K} \cdot \textit{K}$ matrix, filled with \textit{K} distinct symbols, each symbol appearing once in each column and once in each row.
\end{definition}
\begin{definition}\label{orthogonal}
Two distinct \textit{K} $\cdot$ \textit{K} Latin squares E = ($e_{i,j}$) and F = ($f_{i,j}$), so that $e_{i,j}$ and $f_{i,j}$ $\in$ \{1,2, $\dots$ \textit{K}\}, are said to be orthogonal, if the \textit{$K^{2}$} ordered pairs ($e_{i,j},f_{i,j}$) are all different from each other. More generally, the set \textit{OLS}=\{$E_{1}, E_{2}, E_{3},\dots,E_{r}$\} of distinct Latin squares \textit{E} is said to be orthogonal, if every pair in \textit{OLS} is orthogonal.
\end{definition}
\begin{definition}
An orthogonal set of Latin squares of order \textit{K} is of size of (\textit{K-1}), i.e., the number of Latin squares in the orthogonal family is (\textit{K-1}), is called a complete set \cite{key52, key53}.
\end{definition}
\begin{definition}
An $M \cdot K$ Latin rectangle is a $M \cdot K$ matrix G, filled with symbols $a_{ij}$ $\in$ $\{1,2,\dots,K\}$, such that each row and each column contains only distinct symbols.
\end{definition}

To illustrate, E and F are clearly orthogonal Latin squares of order 4, and when superimposed (E $\bowtie$ F), where no two ordered pairs are similar as shown in the Latin square E$ \bowtie$ F.
\begingroup\makeatletter\def\f@size{8}\check@mathfonts
\begin{center}
$E = 
\begin{bmatrix}
   1&2&3&4 \\
   2&3&4&1 \\
   3&4&1&2 \\
   4&1&2&3
\end{bmatrix}$ $F = 
\begin{bmatrix}
   4&1&2&3 \\
   1&2&3&4 \\
   2&3&4&1 \\
   3&4&1&2
\end{bmatrix}$ $J = 
\begin{bmatrix}
   3&4&1&2 \\
   4&1&2&3 \\
   1&2&3&4 \\
   2&3&4&1
\end{bmatrix}$
\end{center}
\begin{center}
$E\bowtie F = 
\begin{bmatrix}
   1,4&2,1&3,2&4,3\\
   2,1&3,2&4,3&1,4\\
   3,2&4,3&1,4&2,1\\
   4,3&1,4&2,1&3,2
\end{bmatrix}$
\end{center}
\endgroup

Throughout this paper, we denote the combination of \say{channel and time-slot assignment} by a symbol in a Latin rectangle.
\section{Interference Mitigation Using Latin Rectangles}
\textit{DAIL} exploits the properties of Latin squares in order to reduce the probability of collision while enabling autonomous scheduling of the medium access. \textit{DAIL} described in detail in the balance of this section.
\subsection{Detailed Algorithm}
To mitigate interference, \textit{DAIL} opts to exploit the availability of multiple channels and allows the individual \textit{WBAN}s to hop among the channels in a pattern that is predictable to the sensors of the same \textit{WBAN} and random to the other coexisting \textit{WBAN}s. To achieve that \textit{DAIL} employs Latin squares as the underlying scheme for channel and time-slot allocation to sensors. Basically, if a \textit{WBAN} picks one Latin square from an orthogonal set, there will be no shared combination of channel and time-slot among the coexisting Latins. According to \textit{\textbf{theorem \ref{theo1}}}, the number of \textit{WBAN}s using orthogonal Latin squares is upper bounded by \textit{K-1}.
\begin{theorem}\label{theo1}
If there is an orthogonal family of r Latin squares of order K, then $r\leq K-1$ \cite{key52}.
\end{theorem}
For detailed proof of \textbf{theorem \ref{theo1}}, refer to \cite{key52, key300}.

The Latin size will depend on the largest among the number of time-slots sensors need, denoted by \textit{K}, and number of channels, \textit{M}. However, the IEEE standard \cite{key26} limits the number of channels which constitutes the rows in the Latin square to \textit{16}, no more than \textit{16} transmissions can be scheduled. To overcome such a limitation, \textit{DAIL} employs Latin rectangles instead, i.e., does not restrict the value of \textit{M} and hence supports \textit{$K > M$}. Thus, regardless whether there is a crowd of patients in a hospital hall or a single patient is sitting in home, each \textit{WBAN}’s \textit{Crd} will autonomously pick a \textit{$M \cdot K$} Latin rectangle orthogonal to potentially coexisting \textit{WBAN}s, i.e., no two or more \textit{Crd}s pick the same Latin rectangle. Then, the \textit{Crd} assigns a single symbol from the set \textit{\{{1,2,\dots,K}\}} to each sensor within its \textit{WBAN}. Afterwards, each sensor determines its transmission pattern, i.e., its channel and time-slot in every superframe according to the position hopping of that symbol in the Latin rectangle. 

\begin{figure}
  \centering
        \includegraphics[width=0.25\textwidth, height=0.14\textheight]{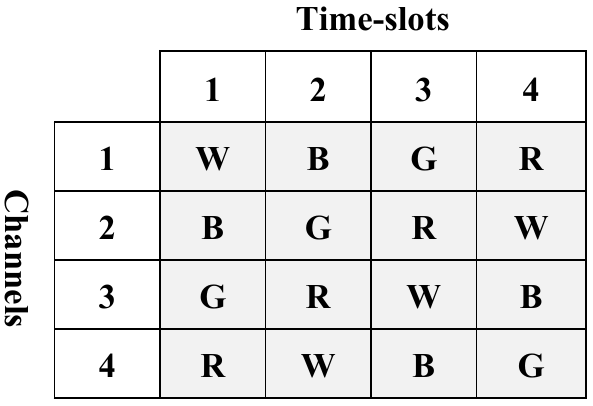}
\caption{A $4 \cdot 4$ channel to time-slot assignment Latin square}
\label{colors}
\end{figure}

The orthogonality property of Latin rectangles avoids inter-\textit{WBAN} interference by allowing each sensor \textit{SR}, to have its unique transmission pattern that does not resemble the pattern of sensors of other \textit{WBAN}s, i.e., they do not share the same position of the symbol, each in its own Latin rectangle and consequently, no other sensor in the network would share the same channel in the same \textit{Slt} with \textit{SR} all the time. For instance, if sensor $RS_u$ is assigned a symbol \say{\textit{B}}, then, its transmission pattern is denoted by \textit{$P_{u}$} = \{$C_{i}Slt_{j}$\}, $\forall i \leq M, \forall j \leq K$, where \textit{C} denotes a channel, will correspond to the positions of \textit{B} in rectangle E as shown in \textbf{figure \ref{colors}}. Then, $P_{u}$ = ($C_{1}Slt_{2}$, $C_{2}Slt_{1}$, $C_{3}Slt_{4}$, $C_{4}Slt_{3}$), i.e., $RS_u$ may transmit through $1^{st}$ channel in $2^{nd}$ \textit{Slt}, $2^{nd}$ channel in $1^{st}$ \textit{Slt}, etc. Therefore, using Latin rectangles, each \textit{Crd} prevents the interference through orthogonal channel to time-slot assignments hopping. 

Generally, \textit{DAIL} makes it highly improbable for two transmissions to collide as we show in Section \rom{5}. Nonetheless, collision may still occur when (i) two \textit{WBAN}s randomly pick the same Latin rectangle, or (ii) more than \textit{16} \textit{WBAN}s coexist in the same area, which means that, the number of \textit{WBAN}s exceeds the number of \textit{ZIGBEE} channels (\textit{16}) in the Latin rectangle. \textit{DAIL} handles these cases by extending the superframe size through increasing the number of columns in the Latin rectangle, i.e., increasing the number of \textit{Slt}s in the Latin. In the next section we determine the setting of superframe size and in Section \rom{5} we analyze how to set \textit{K} per each superframe. \textbf{Algorithm \ref{DAIL}} provides a high level summary of \textit{DAIL}.

\setlength{\textfloatsep}{1pt}
\begin{algorithm}
\footnotesize
\SetKwData{Left}{left}\SetKwData{This}{this}\SetKwData{Up}{up}
\SetKwFunction{Union}{Union}\SetKwFunction{FindCompress}{FindCompress}
\SetKwInOut{Input}{input}\SetKwInOut{Output}{output}

\Input{\textit{N} \textit{WBAN}s, Coordinator \textit{Crd}, \textit{M} ZIGBEE channels, Latin rectangle \textit{R}, frame length \textit{FL}}

\textbf{BEGIN}
 
\quad \textit{FL =  K}\: //\; default setting of the frame length 
 
\quad \textit{\textbf{if}} \textit{N $>$ K} \textit{\textbf{then}}

\quad \quad \quad \textit{FL = N}\: //\; \textit{Crd} increases the number of time-slots in the superframe

\quad\textit{\textbf{endif}}

\quad Each WBAN's \textit{Crd} randomly picks a Latin rectangle \textit{R} of size \textit{M $\cdot$ F}

\textbf{END}

\caption{Proposed \textit{DAIL} Scheme}
\label{DAIL}
\end{algorithm}
\DecMargin{1em}

\subsection{Superframe Size}
While, we consider all \textit{M = 16} channels of \textit{ZIGBEE} available at each \textit{WBAN}, we still need to determine the number of time-slots per each row of Latin rectangle, in other words, the length of each superframe. Each sensor $SR_{i}$, where $i \leq L$, may require \textit{p} \textit{Slt}s to complete its data transmission. For example, for a sensor that samples at a rate of 10 per second, we need 10 \textit{Slt}s in a frame of 1 second. If all sensors have the same requirement, \textit{$p \cdot L$} \textit{Slt}s for \textit{L} sensors are required in each frame. In fact, the frame size depends on two factors, 1) how big the \textit{Slt}, which is based on the protocol in use, and 2) the number of required \textit{Slt}s, which is determined by the different sampling rates of \textit{WBAN} sensors. Generally, the sum of number of samples for all sensors in a time period determines the frame size. However, \textit{DAIL} requires the frame size for all \textit{WBAN}s to be the same so that collision could be better avoided by picking the right value for \textit{K}. Therefore, in \textit{DAIL} the superframe size is determined based on the highest sampling rate. In this case, the number of \textit{Slt}s to be made in the superframe, respectively, in the Latin rectangle is \textit{K} slots, where \textit{K = \textit{p} $\cdot$ \textit{L}}.
\subsection{Illustrative Example}
\begin{figure}
  \centering
        \includegraphics[width=0.3\textwidth]{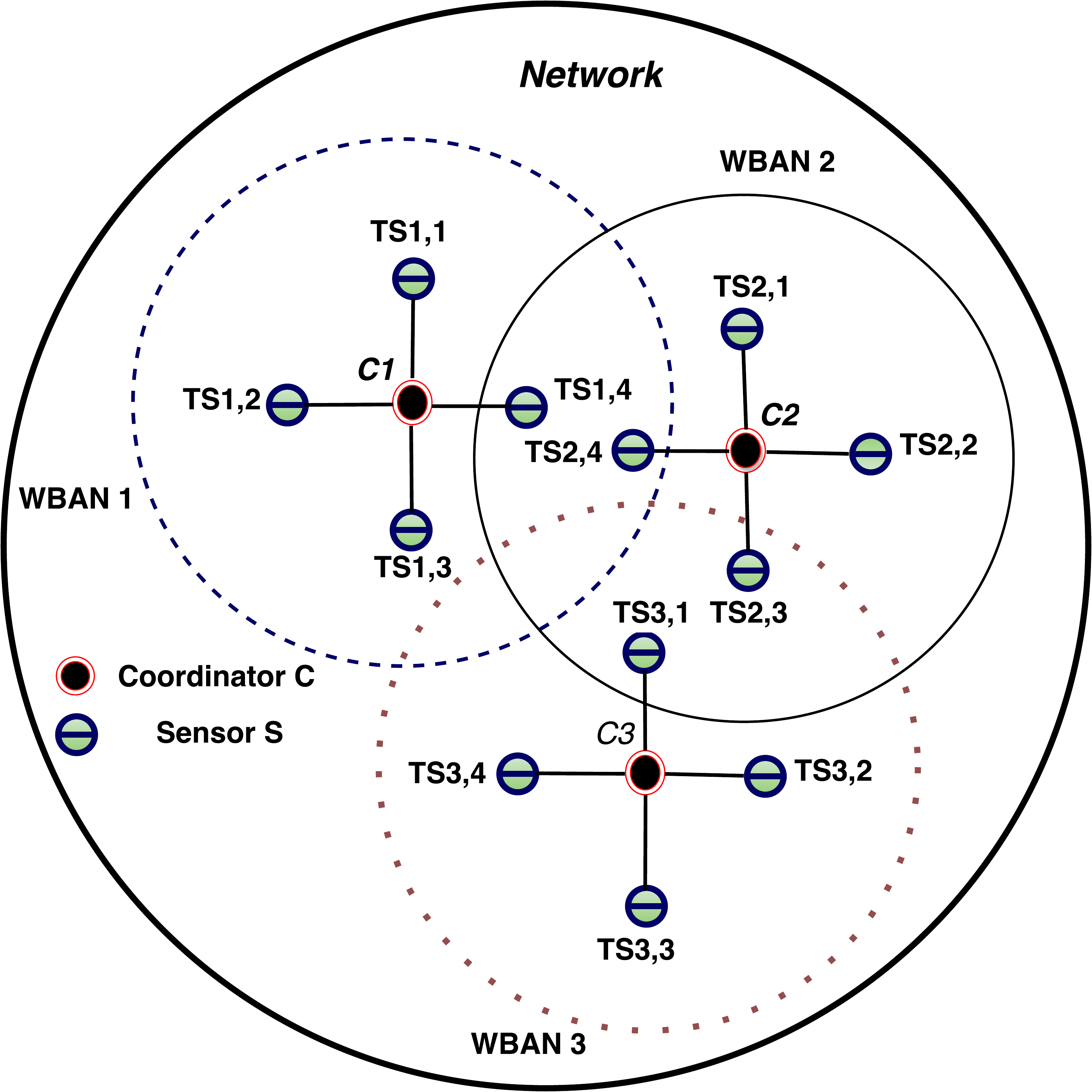}
\caption{A network consisting of three coexisting TDMA-based \textit{WBAN}s}
\label{coexistwbans}
\end{figure}

We illustrate our approach through a scenario of \textit{3} coexisting \textit{WBAN}s, where each circumference represents the interference range as shown in \textbf{figure \ref{coexistwbans}}. Furthermore, each \textit{WBAN} is assigned \textit{M = 4} channels and consists of \textit{L = 4} sensors, in turn, each sensor is assigned a symbol from the set \textit{K = \{1,2,3,4\}$\iff$\{G,B,R,W\}}. Here, we assume that each sensor requires only one \textit{Slt} to transmit its data in each superframe.
Based on this scenario, any pair of sensors are interfering with each other, i.e., they transmit using the same channel in the same time, if both sensors are in the intersection of their corresponding interference ranges. However, as shown in \textbf{figure \ref{coexistwbans}}, $4^{th}$ sensor of $\textit{WBAN}_1$ denoted by $SR_{1,4}$ and $SR_{2,4}$ are interfering, also, $SR_{3,1}$ and $SR_{2,3}$. Therefore, to address this problem, each \textit{WBAN} picks a distinct Latin rectangle from an orthogonal set as follows: $\textit{WBAN}_1$ picks E, $\textit{WBAN}_2$ picks F and $\textit{WBAN}_3$ picks J, where E, F and J are considered as in (\rom{3}-B). Assume 3 sensors, $SR_u$, $SR_v$ and $SR_w$ of $\textit{WBAN}_1$, $\textit{WBAN}_2$ and $\textit{WBAN}_3$ are, respectively, assigned symbols B, R and G in Latin rectangles E, F and J. Thus, the distinct positions of symbol B in E corresponds to the transmission pattern $P_u$ in $\textit{WBAN}_1$'s superframe, similarly for $P_v$ and $P_w$ in $\textit{WBAN}_2$ and $\textit{WBAN}_3$, respectively. However, \textit{B=2 in E, R=3 in F and G=1 in J}, therefore, the transmission patterns for  $P_u$, $P_v$ and $P_w$ are, respectively, represented by B, R and G symbols of the matrix shown in \textbf{figure \ref{colors}}. As clearly seen in this figure that $SR_u$, $SR_v$ and $SR_w$ neither share the same channel nor the same \textit{Slt}, i.e., no collision occurs at all.
\section{Mathematical Analysis}
Although Latin rectangles diminishes inter-\textit{WBAN}s interference, there are still some possibilities for collisions as pointed out in Section \rom{4}. Therefore, it is necessary to analyze the probability of collisions among the sensors of different \textit{WBAN}s. In this section we opt to analyze the performance of \textit{DAIL} mathematically. We consider a multichannel \textit{TDMA}-based network, where superframes are constructed as an $M \cdot K$ matrix, where within each superframe, each sensor may be assigned \textit{\textit{M}} \textit{Slt}s to transmit its data according to a unique channel to \textit{Slt} assignment pattern. These patterns are generated from the orthogonal family of $M \cdot K$ Latin rectangles. However, all sensors of each \textit{WBAN} share one common $M \cdot K$ Latin rectangle, where, the pattern of each sensor corresponds to a single symbol pattern in that rectangle, as shown in \textbf{figure \ref{colors}}. 
\subsection{Interference Bound}
In this subsection we opt to determine the worst-case collision pattern for the individual sensor.
\begin{definition}
Let E and F be two orthogonal $M \cdot K$ Latin rectangles. Symbol e from E is assigned to $SR_u$, and symbol f from F is assigned to $SR_v$. Then, there exists a collision at the $j^{th}$ \textit{Slt} on $i^{th}$ channel for $SR_u$ and $SR_v$, if the ordering (e,f) of both rectangles appears at $i^{th}$ row, $j^{th}$ column, which means $[E_{i,j}] = e$ and $[F_{i,j}] = f$.
\end{definition}
\begin{theorem}\label{coll}
If two sensors are assigned two distinct symbols in the same Latin rectangle, there will be no collision among their transmissions. If they are assigned symbols from two distinct orthogonal Latin rectangles, then, they will face at most one collision in every superframe.
\end{theorem}
\textbf{Proof:} From the definition of Latin rectangles, because every symbol occurs exactly one time in each row and exactly one time in each column, any two \textit{Slt} assignment patterns constructed from the same Latin rectangle will not have any overlap in their patterns and so they will not have any collision with each other. Based on \textbf{defintion \ref{orthogonal}}, hence, the ordering (e,f) for any pair of orthogonal Latin rectangles, where, e and f $\in$ \textit{\{1,2,\dots,K\}}, can only appears one time, which means that, these sensors will only have one opportunity of collision.
\begin{theorem}
In a crowded network of \textit{N} \textit{WBAN}s, each sensor has a channel to \textit{Slt} transmission pattern corresponding to a symbol pattern chosen from one of the $K^{th}$ set of orthogonal Latin rectangles. Let us consider a sensor denoted by \say{s} surrounded by maximum number of \textit{Q} \textit{WBAN}s, i.e., \textit{Q} sensors from other \textit{WBAN}s, which means, \textit{Q} sensors may coexist in the communication range of \textit{s}. Then, \textit{s} may experience at most Q collisions. Additionally, sensor \textit{s} may face a minimal number of collisions which equal to \textit{max((Q-K+1),0)}.
\end{theorem}

\textbf{Proof:} Based on \textbf{theorem \ref{coll}}, each neighboring sensor can create at most one collision to \textit{s}. In the worst case, all \textit{Q} sensors are within the range of communication of \textit{s}. The transmissions patterns of \textit{Q} sensors are constructed from Latin rectangles that are different from the Latin rectangle utilized by \textit{s}. Subsequently, the maximum number of possible collisions experienced by \textit{s} is \textit{Q}. Now, to count the minimal number of collisions for \textit{s}, it is required to find the maximum number of sensors that construct their transmission patterns from the same Latin rectangle, which is \textit{K}, i.e., \textit{K} sensors will have no collision according to \textbf{theorem \ref{coll}}. Also, \textbf{theorem \ref{coll}} proves that there exists at most one collision for each pair of sensors constructing their transmission patterns from two different orthogonal Latin rectangles. Therefore, each of the remaining sensors \textit{(Q-K+1)} will cause one collision to \textit{s} because they belong to different orthogonal Latin rectangles. As a result, the minimum number of collisions for sensor \textit{s} surrounded by \textit{Q} sensors is equal to \textit{max((Q-K+1),0)}.
\subsection{Collision Probability}
We consider a sensor $SR_i$ of $\textit{WBAN}_i$ is surrounded by \textit{Q} interfering sensors $v_j$ of different coexisting $\textit{WBAN}_j$ in the vicinity, where $\textit{j=1,2,\dots,Q}$ and $i\neq j$. For simplicity, we assume, each sensor transmits one data packet in each $Slt$. However, sensor $SR_i$ successfully transmits its data packet in $Slt_{i}$ and on channel $C_{i}$ to the \textit{$Crd_{i}$}, \textit{iff}, none of the \textit{Q} neighbors transmits its data packet using the same $Slt$ on the same channel as sensor $SR_i$. Let \textit{X} denotes the random variable representing the number of sensors that are transmitting their data packets in the same $Slt$ as sensor $SR_i$, if x packets are transmitted in the the same $Slt$ as $SR_i$. Then, the probability of event \textit{X} is defined by \textbf{equation} (\ref{eq20}) below.

\small
\begin{multline}\label{eq20}
         Pr\left(X=x\right)=C_{x}^{Q+1}\cdot \omega^{x}\cdot (1-\omega)^{Q-x}\cdot \left(min(M,K)/K\right)^{x}\\
         \forall\: x\: \leq\: Q
\end{multline}
\normalsize
Where $\omega$ is the use factor, defined as the ratio of the time that a sensor is in use to the total time that it could be in use. Now, suppose \textit{Y} sensors out of \textit{X} sensors schedule their transmissions according to the same Latin rectangle as sensor $SR_i$, i.e. y out of x sensors select symbol patterns from the same Latin rectangle as $SR_i$.

\small
\begin{multline}\label{eq21}
        Pr\left(Y=y \mid X=x\right)=\left(C_{y}^{K+1}\cdot C_{x-y}^{Z-K}\right)/C_{x}^{Z-1}\\
                                   \forall x \leq Q \: \& \:  \forall \: y\: \leq \: x
\end{multline}
\normalsize
Where \textit{Z = K$\cdot$ m} is the total number of symbol patterns in the orthogonal Latin rectangle family. However, these \textit{Y} sensors will not impose any collision with $SR_i$'s transmission, since they (\textit{Y} sensors) use the same Latin rectangle as $SR_i$. On the other hand, \textit{X-Y} sensors may collide with the transmission from sensor $SR_i$ to the \textit{Crd} on the same channel, then the conditional probability of transmission collision is denoted by (\textit{collTx}) and defined by \textbf{equation} (\ref{eq22}) below.

\small 
\begin{multline}\label{eq22}
Pr(collTx\mid\: Y=y\: \& \: X=x)\\
=1-Pr(succTx\: \mid\: Y=y\: \& \: X=x)\\
=1-\left((min(M,K)-1)/min(M,K)\right)^{x-y}
\end{multline}
\normalsize
Where \textit{min(M,K)} represents the number of transmission \textit{Slts} for each sensor in each superframe. Then, the probability of a successful data packet transmission from sensor $SR_i$ to the \textit{Crd} is denoted by $\lambda$ as follows:

\small 
\begin{equation}
\begin{split}
\lambda=&\sum_{x=0}^{Q}\sum_{y=0}^{x}Pr(Y=y,X=x)\\
        &\cdot Pr(succTx\: \mid Y=y\: \&\: X=x)\\
        &=\sum_{x=0}^{Q}\sum_{y=0}^{x}Pr(Y=y\: \mid\: X=x)\cdot Pr(X=x)\\
        &\cdot Pr(succTx\: \mid\: Y=y\: \&\: X=x)\\
        &=\sum_{x=0}^{Q}\sum_{y=0}^{x}(C_{x}^{Q} C_{y}^{K-1} C_{x-y}^{Z-K})/(C_{x}^{Z-1})\\
        &\cdot \omega^x \cdot (1-\omega)^{Q-x}\cdot \left(min(M,K)/K\right)^{x}\\
        &\cdot \left((min(M,K)-1)/min(M,K)\right)^{x-y} 
\end{split}
\end{equation}\label{eq90}
\normalsize
\/*
\subsection{Throughput Analysis}
Let the size of the orthogonal family of $K^{th}$ order Latin squares is \textit{m = K-1} and the transmission pattern of each sensor is determined by one of the \textit{$K^{2}$} distinct symbol patterns in the $K\cdot K$ Latin square. When $K > M$, each $K\cdot K$ Latin square can be cut into $M \cdot K$ Latin rectangle. To assure that every sensor has unique transmission pattern according to these Latin rectangles, ($K \cdot m \geq N$) must be satisfied, where \textit{N} is the number of \textit{WBAN}s. Furthermore, it has been proven in \textbf{theorem \ref{coll}} that the number of collisions (\textit{\# colls}) in each superframe for any two sensors is either one or zero. Assuming the maximum number of neighbors to \textit{SR} is still \textit{Q}, then, each sensor will be assigned \textit{min(M,K)} transmission \textit{Slts} in each superframe denoted by \textit{SF}. We denote by \textit{TS} the number of successful transmissions for each sensor, \textit{$TS_{min}$} and \textit{$TS_{max}$} are the lower and the upper bounds of \textit{TS}, respectively, when \textbf{equation} (\ref{eq2}) holds, every sensor will have its throughput in \textbf{equation} (\ref{eq4}) and \textbf{equation} (\ref{eq5}) as follows:

\small 
\begin{equation}\label{eq2}
K \geq TS_{max} \geq TS \geq TS_{min} > 0 
\end{equation}
\normalsize

\small 
\begin{equation}\label{eq3}
TS = min(M,K)-(\#\: colls\: per\: SF)
\end{equation}
\normalsize

\small 
\begin{equation}\label{eq4}
TS_{max} =
  \begin{cases}
     K-max(Q-K+1,0)\:  &\: if\: K\: \leq\: M\\
     M-max(Q-K+1,0)\:  &\: if\: K\: >\: M\\ 
  \end{cases}
\end{equation}
\normalsize

\small 
\begin{equation}\label{eq5}
 TS_{min} =
  \begin{cases}
   K-Q\: &\: if\: K\: \leq\: M\\
   M-Q\: &\: if\: K\: >\: M\\
  \end{cases}
\end{equation}
\normalsize
Therefore, to assure that every sensor has a minimal throughput, \textit{K} should be greater than \textit{Q} when \textit{K} $\leq$ \textit{M}, or \textit{M} should be greater than Q when $K > M$. In order to evaluate the performance of our approach, the best and the lowest throughput, respectively, denoted by \textit{$T_{max}$} and \textit{$T_{min}$} are defined in \textbf{equation} (\ref{eq7}) and \textbf{equation} (\ref{eq100}).
\begin{definition}
\textit{$T_{max}$} (resp. \textit{$T_{min}$}) is defined as the ratio of the maximal (resp. minimal) number of successful transmissions in each \textit{SF} to its length denoted by \textit{FL}
\end{definition}

\small 
\begin{equation}\label{eq7}
T_{max}=TS_{max}/FL,\: FL = K
\end{equation}
\normalsize

\small 
\begin{equation}\label{eq100}
T_{min}=T_{min}=TS_{min}/FL),\: FL = K
\end{equation}
\normalsize
\begin{theorem}\label{theo2}
For given \textit{Q}, \textit{N} and \textit{M}, the maximal nonzero upper and lower bounds of throughput \textit{T} are as follows:
\end{theorem}

\small
\begin{equation}\label{eq8}
1\: \geq\: T\: \geq\: 1-(Q/M)\: , if\: K\: \leq\: M
\end{equation}
\normalsize

\small 
\begin{multline}\label{eq9}
M/max(M,\floor{N/m})\geq T\geq (M-Q)/max(M,\floor{N/m}),\\
\: if\: K>M
\end{multline}
\normalsize
\textbf{Proof:} When $K \leq M$, based on \textbf{equation} (\ref{eq7}), the upper and lower bounds of \textit{T} are as follows:

\small 
\begin{multline}\label{eq12}
T_{max}=TS_{max}/FL=\left(K-max(Q+1-K,0)\right)/K\\
1-\left(max(Q-K+1,0)/K\right)
\end{multline}
\normalsize

\small 
\begin{equation}\label{eq13}
T_{min}=TS_{min}/FL=(K-Q)/K=1-Q/K
\end{equation}
\normalsize
We can deduce from \textbf{equation} (\ref{eq12}) and \textbf{equation} (\ref{eq13}) that the upper and lower bounds of \textit{T} will increase with \textit{K}. Thus, to ensure that the minimal throughput is greater than zero and every sensor has a unique transmission pattern, then, this inequality; $Q<K<\ceil{N/m}$ must be satisfied. Also, we can have, \textit{$max(Q+1-K,0)=0$} and \textit{$\ceil{N/m}\leq K\leq M$}. Therefore, when \textit{K = M}, the maximal upper and lower bounds of the throughput are shown in \textbf{equation} (\ref{eq14}) and \textbf{equation} (\ref{eq15}) below.

\small 
\begin{equation}\label{eq14}
T_{max} = 1 \: and \: T_{min} = 1-Q/M
\end{equation}
\begin{equation}\label{eq15}
T_{min} = 1-Q/M
\end{equation}
\normalsize
Similarly, if \textit{$K>M$}, the bounds of \textit{T} are shown in \textbf{equation} (\ref{eq16}) and \textbf{equation} (\ref{eq17}) below.

\small 
\begin{multline}\label{eq16}
\begin{split}
 T_{max}=&TS_{max}/FL\\
& =\left(M-max(Q+1-K,0)\right)/K=M/K
 \end{split}
\end{multline}
\begin{equation}\label{eq17}
 T_{min}=TS_{min}/FL=(M-Q)/K
\end{equation}
\normalsize
However, these bounds decrease when \textit{K} increases. So, when \textit{$K > \ceil{N/m}$} and \textit{$K > M$} are combined, then, \textit{$K>max(M,\ceil{N/m})$} is true, and so the maximal upper and lower bounds of \textit{T} are as in \textbf{equation} (\ref{eq18}) and \textbf{equation} (\ref{eq19}) below. 

\small 
\begin{equation}\label{eq18}
 T_{max}=M/max(M,\ceil{N/m})
\end{equation}
\begin{equation}\label{eq19}
 T_{min}=(M-Q)/max(M,\ceil{N/m})
\end{equation}
\normalsize
When \textit{$K=max(M,\ceil{N/m})$}. In \textbf{theorem \ref{theo2}}, when \textit{$M \geq K$} corresponds to the number of available channels is greater than the number of transmission \textit{Slt}s assigned to a sensor in a \textit{WBAN}, however, the minimal throughput \textit{$T_{min}$} can be maximized when we choose \textit{K} equals to the maximal number of available channels, which is limited to \textit{M} in our case, and so, \textit{$M < K$}. Therefore, the bounds of the throughput will be impacted by the size of the Latin rectangle family \textit{m}.
*/
\begin{figure*}
\begin{minipage}[b]{.3075\textwidth}
\centering
\includegraphics[width=1\textwidth, height=0.2\textheight]{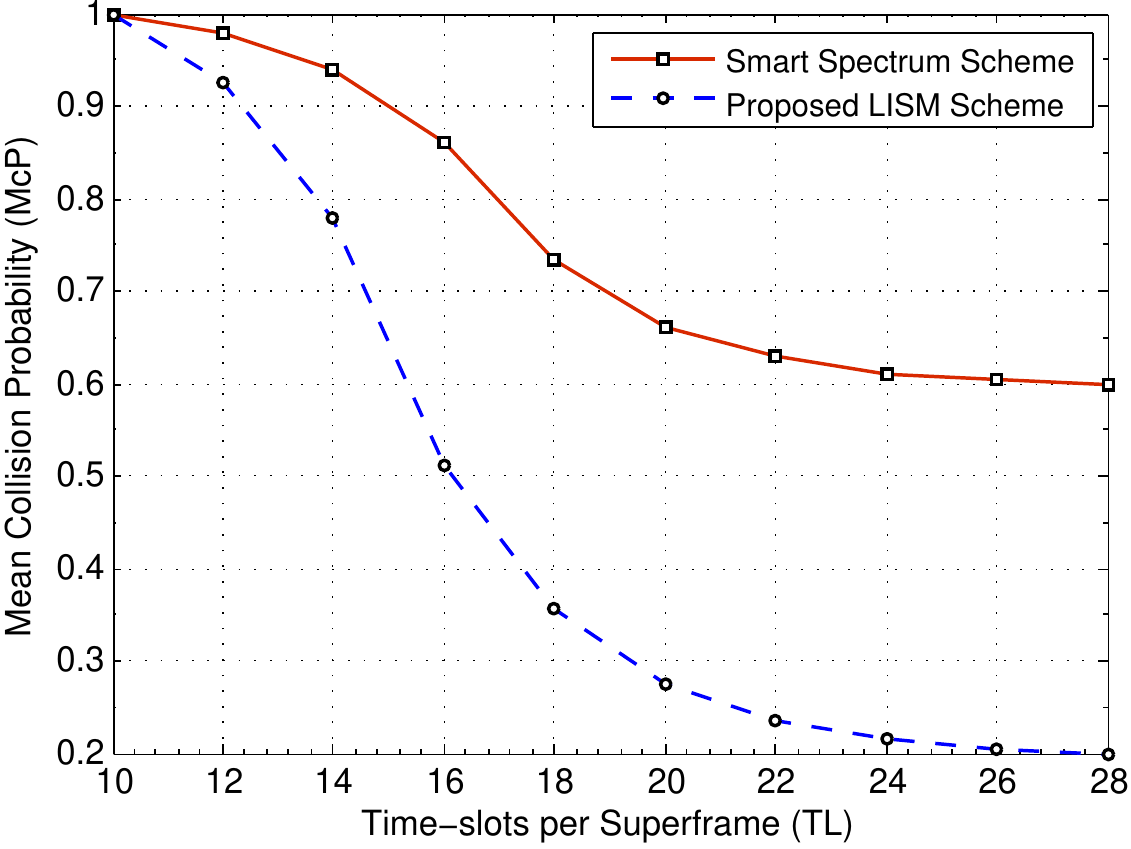}
\caption{\textit{McP} versus \# of slots per superframe (\textit{TL})}
\label{slots}
\end{minipage}\qquad
\begin{minipage}[b]{.3075\textwidth}
\centering
\includegraphics[width=1\textwidth, height=0.2\textheight]{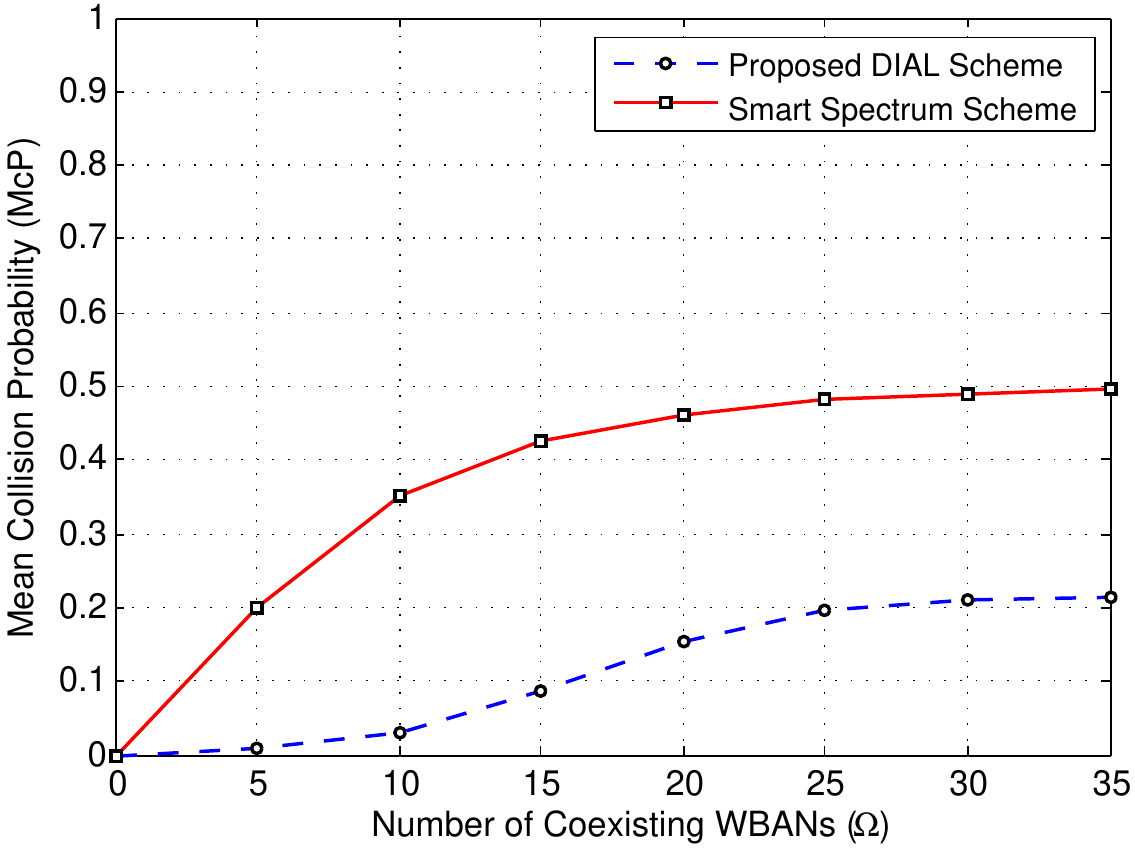}
\caption{\textit{McP} versus \# of coexisting \textit{WBAN}s ($\Omega$)}
\label{wbans}
\end{minipage}\qquad
\begin{minipage}[b]{.3075\textwidth}
\centering
        \includegraphics[width=1\textwidth, height=0.2\textheight]{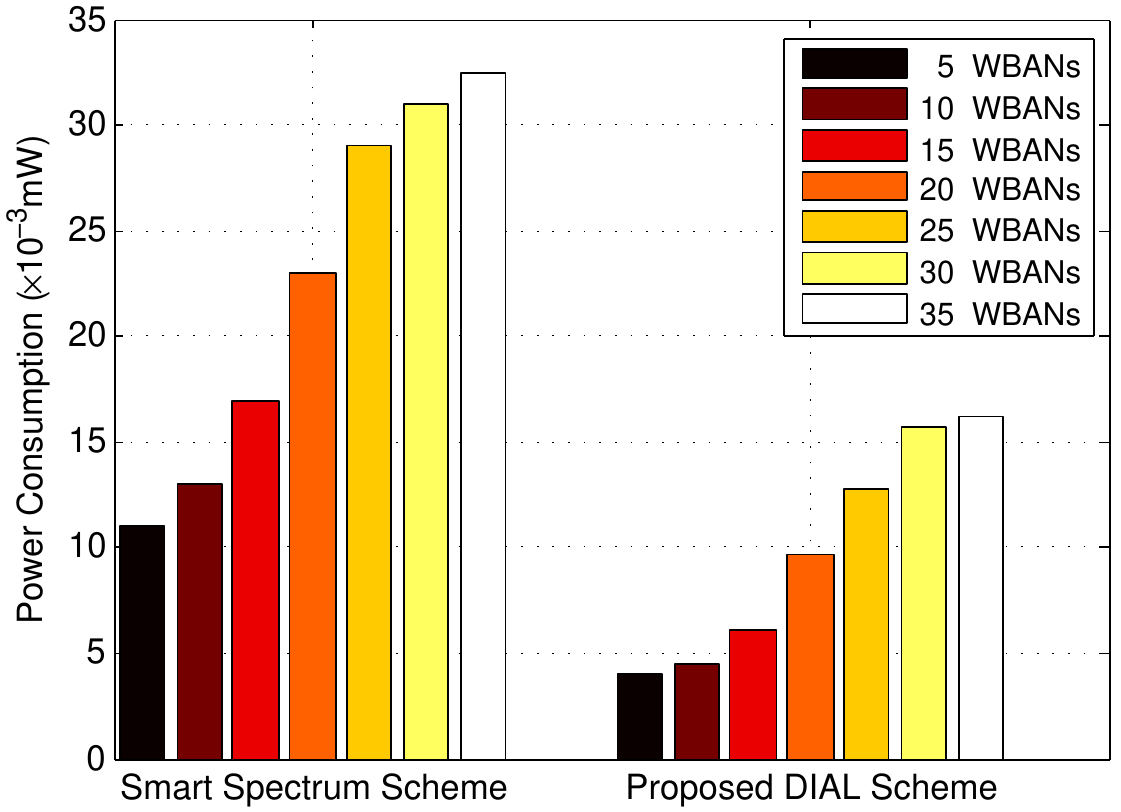}
\caption{Mean power consumption (\textit{PC}) versus ($\Omega$)}
\label{energy}
\end{minipage}\qquad
\end{figure*}

\section{Performance Evaluation}
We have performed simulation experiments to validate the theoretical results and evaluate the performance of the proposed \textit{DAIL} scheme. In this section, we compare the performance of \textit{DAIL} with the smart spectrum allocation scheme, denoted by \textit{SMS} \cite{key16}, which assigns orthogonal channels to interfering sensors belonging to each pair of coexisting \textit{WBAN}s. The simulation parameters are provided in \textbf{table \ref{parm}}.
\begin{table}
\centering
\caption{Simulation Setup $\&$ Parameters}
\label{parm}
\begin{tabular}{lllll}
\noalign{\smallskip}\hline
\hline\noalign{\smallskip}
&\textbf{Exp. 1} & \textbf{Exp. 2} &\textbf{Exp. 3}&\textbf{Exp. 4}\\
Sensor TxPower(dBm) & -10 &-10 &-10&-10\\
\# \textit{Crd}s/\textit{WBAN} & 1 &1 &1 & 1 \\
\# Sensors/\textit{WBAN} &12 &12 & 12&12\\
\# \textit{WBAN}s/Network &Var & 30 & Var&Var\\
\# Slots/Superframe  &12 &12 &12& 12\\
Latin Rectangle Size&$16 \cdot 12$ &$16 \cdot Var$ &$16 \cdot 12$& $16 \cdot 12$\\
\noalign{\smallskip}\hline
\hline\noalign{\smallskip}
\end{tabular}
\end{table}
\subsection{Collision Probability}
In experiment 1, the mean collision probability denoted by \textit{McP} versus the number of coexisting \textit{WBAN}s ($\Omega$) for \textit{DAIL} and that for \textit{SMS} are compared in \textbf{figure \ref{wbans}}. As can be clearly seen in the figure, \textit{DAIL} provides a much lower \textit{McP} because of the combined channel and \textit{Slt} hopping. It is observed from this figure that \textit{McP} of \textit{DAIL} is very low when $\Omega \leq 12$ due to the large number of channel and \textit{Slt} combinations. When $12 < \Omega \leq 25$, \textit{McP} significantly increases due to the growth in the number of sensors which makes it possible for two or more sensors to be assigned the same channel in the same time-slot. However, when $\Omega$ exceeds 25, \textit{McP} increases very slightly and eventually stabilizes at $21 \cdot 10^{-2}$ because of the maximal number of collisions is attained by each \textit{WBAN}. In \textit{SMS}, \textit{McP} significantly increases when $0 < \Omega \leq 18$, i.e., the number of channels and the number of \textit{WBAN}s are similar. Then, \textit{McP} slightly increases until it stabilizes at $5 \cdot 10^{-1}$ when $18 < \Omega \leq 35$ since the interference attains its maximum and all channels are already assigned. \textit{McP} significantly grows for as long as the number of channels is smaller than $\Omega$. However, when $\Omega$ exceeds \textit{16}, \textit{McP} tends to stabilize at $5 \cdot 10^{-1}$.

Meanwhile, experiment 2 studies the effect of the number of \textit{Slt}s per a superframe denoted by \textit{TL} on \textit{McP}. As can be clearly seen in \textbf{figure \ref{slots}}, \textit{DAIL} always achieves lower collision probability than \textit{SMS} for all \textit{TL} values. In \textit{DAIL}, \textit{McP} significantly decreases as \textit{TL} increases from $10$ to $28$, where increasing \textit{TL} is similar to enlarging the size of the Latin rectangle. Therefore, a larger number of channel and \textit{Slt} combinations allows distinct sensors to not pick the same channel in the same \textit{Slt}, which decreases the chances of collisions among them. However, \textit{SMS} depends only on the 16 channels to mitigate interference, and the channel assigned to a sensor stays the same for all the time. Thus, a high \textit{McP} is expected due to the larger number of interfering sensors than the available channels. Moreover, a sensor has \textit{16} possibilities in \textit{SMS}, while it has $16 \cdot framesize$ different possibilities in \textit{DAIL} to mitigate the interference, which explains the large difference in \textit{McP} amongst two schemes.

\subsection{\textit{WBAN} Power Consumption}
In experiment 3, the power consumption of each \textit{WBAN} denoted by \textit{PC} versus the number of coexisting \textit{WBAN}s ($\Omega$) for \textit{DAIL} and \textit{SMS} are compared. \textbf{Figure \ref{energy}} shows that \textit{PC} for \textit{DAIL} is always lower than that of \textit{SMS} for all values of $\Omega$. Such distinct performance for \textit{DAIL} is mainly due to the reduced collisions that lead to fewer retransmissions and consequently lower power consumption. For \textit{DAIL}, the figure shows that \textit{PC} slightly increases when $\Omega \leq 10$, i.e., there is a larger number of channel and \textit{Slt} combinations than the interfering sensor pairs which lowers the number of collisions among sensors and hence the power consumption is decreased. \textit{PC} significantly increases when $10 < \Omega \leq 30$ due to the large number of sensors competing for the same channel in the same \textit{Slt}s which results in more collisions and hence more power consumption. When $\Omega$ exceeds 30, the power consumption increases slightly to attain the maximum of $16.5 \cdot 10^{-3}mW$. However, in \textit{SMS}, \textit{PC} is high due to the collisions resulting from the large number of sensors that compete for the available channels (\textit{16 channels}).
\/*
\subsection{\textit{WBAN} Throughput}
In experiment 4, the mean successful packets received at each \textit{WBAN} denoted by \textit{MsPR} versus the number of coexisting \textit{WBAN}s ($\Omega$) for \textit{DAIL} and \textit{SMS} are compared. \textbf{Figure \ref{thro}} shows that \textit{DAIL} always achieves higher \textit{MsPR} than \textit{SMS} for all values of $\Omega$. Such improvement in \textit{DAIL} performance is mainly because of the reduced collisions, which boosts the number of data packets that are successfully received in a period of time. For \textit{DAIL}, the figure shows that \textit{MsPR} significantly increases when $\Omega \leq 10$, i.e., there is a larger number of channel and \textit{Slt} combinations than the interfering sensor pairs, which lowers the number of collisions among sensors and hence the number of successfully received packets for each \textit{WBAN} grows. \textit{MsPR} slightly increases when $10 < \Omega \leq 25$ due to the large number of sensors competing for the same channel in the same \textit{Slt}s which results in more collisions. When $\Omega$ exceeds 25, \textit{MsPR} increases very slightly to stabilize at $198$ because of the medium contention reaches its maximum level. However, in \textit{SMS}, \textit{MsPR} significantly increases as $\Omega$ grows for as long as $\Omega \leq 15$ due to the availability of a larger the number of channels than the number of interfering sensors. When $15 < \Omega \leq 25$, a low \textit{MsPR} observed since many sensors compete for the number of available channels (\textit{16 channels}). When $\Omega$ exceeds 25, \textit{MsPR} increases very slightly and eventually stabilize at 67, where each \textit{WBAN} experiences the highest level of medium access contention.
*/
\section{Conclusions}
In this paper, we have presented \textit{DAIL}, a distributed  TDMA-based interference avoidance scheme for coexisting \textit{WBAN}s based on the properties of Latin squares. \textit{DAIL} combines the channel and time-slot hopping to lower the probability of collisions among transmission of sensors in different coexisting \textit{WBAN}s. Accordingly, each distinct \textit{WBAN}'s \textit{Crd} autonomously picks an orthogonal Latin rectangle and assigns its individual sensors unique transmission patterns. Compared with most existing algorithms, \textit{DAIL} has low complexity and does not require any inter-\textit{WBAN} coordination. We have analyzed the expected collision probability. Simulation results show that \textit{DAIL} outperforms other competing schemes.

\end{document}